\documentclass{ws-procs975x65}

\newcommand{\be}{\begin{eqnarray}}
\newcommand{\ee}{\end{eqnarray}}

\begin{document}

\title{EQUATION OF STATE IN THE INNER CRUST OF NEUTRON STARS: DISCUSSION OF THE UNBOUND NEUTRONS STATES}

\author{J. MARGUERON and N. VAN GIAI}

\address{Institut de Physique Nucl\'eaire, Universit\'e Paris-Sud \\
F-91406 Orsay CEDEX, France}

\author{N. SANDULESCU}

\address{Institute of Physics and Nuclear Engineering,\\
R-76900 Bucharest, Romania}

\begin{abstract}
In this paper, we calculate the stable Wigner-Seitz (W-S) cells in the inner
crust of neutron stars and we discuss the nuclear shell effects. 
A distinction is done between the shell effects due to the bound states 
and those induced by the unbound states, which are shown to be spurious. 
We then estimate the effects of the spurious shells on the total energy and
decompose it into a smooth and a residual part. We propose a
correction to the Hartree-Fock binding energy in Wigner-Seitz cell
(HF-WS).
\end{abstract}

\keywords{Equation of state, self-consistent mean-field model, pairing correlations.}

\bodymatter

\section{Introduction}

From the very first nuclear models based on extrapolations of a
mass formula~\cite{bet70}, several models have been developed to
provide the equation of state in the inner crust, like the
BBP~\cite{bbp} and BPS~\cite{bps} semi-classical models or fully
quantal self-consistent mean-field
models~\cite{negele,mag,mon04,bar97,baldo,gog07}. The latter
models are based on the Wigner-Seitz (W-S) approximation and the
continuum states are discretized in a box (spherical or not). In
semi-classical models, a clear limitation is coming from the
absence of nuclear shell effects which is known to play an
important role in determining the composition of the ground state
prior to the neutron drip~\cite{negele}, in the inner
crust~\cite{mon04} as well as in the transition zone between the
crust and the core~\cite{bul02}. Since the seminal work of Negele
\& Vautherin~\cite{negele}, many efforts have been invested in the
developpement of self-consistent mean-field modelization of the
inner crust, by introducing deformation of the nuclear cluster in
the high density part of the crust~\cite{mag,gog07}, or pairing
correlations~\cite{bar97,sandulescu2004,baldo}, or more recently
by improving of the lattice description within the band
theory~\cite{chamel}. Band theory takes into account the proper
symmetries of the system, but the equations are numerically very
complicated to solve and have not yet been solved in a
self-consistent framework. Instead, it has been coupled to a
self-consistent mean-field model at the W-S
approximation~\cite{cha07}. It has then been shown that the W-S
approximation is justified if the temperature is larger than about
100~keV, or if the quantity of interest averages the density of
states on a typical scale of about 100~keV around the Fermi
surface. The typical scale which has been found, 100~keV, is in
fact related to the average inter-distance energy between the
unbound states which are discretized in the W-S box. In this
paper, we clarify the effect of the discretization of the unbound
states on the ground state energy as well as the role of pairing
correlations. Consequences for the equation of state are
presented.

\section{Equation of state in the inner crust}
\label{jm:sec1}

\begin{table}[t]
\setlength{\tabcolsep}{0.06in}
\renewcommand{\arraystretch}{0.9}
\tbl{Comparison of the proton number $Z$ and the radius of the W-S cell
  $R_{WS}$ obtained in different calculations. See the text for more
  explanations.}
{\begin{tabular}{@{}cccccccccc@{}}
\toprule
$N_{zone}$ & $k_F$ & \multicolumn{4}{c}{Z} & \multicolumn{4}{c}{$R_{WS}$ ( in fm) } \\
 & fm$^{-1}$ & N-V\cite{negele} & HF-BCS$^{\text a}$ & HF & HFB & N-V\cite{negele} & HF-BCS$^{\text a}$ & HF & HFB \\
\colrule
 2 & 1.12 & 40 & 20 & 22  & 22 & 19.6 & 15 & 16.6 & 16.6 \\
 3 & 0.84 & 50 & 36 & 22  & 20 & 27.6 & 24 & 20.6 & 20.6 \\
 4 & 0.64 & 50 & 56 & 24  & 24 & 33.2 & 36 & 26.2 & 26.6 \\
 5 & 0.55 & 50 & 58 & 28  & 20 & 35.8 & 40 & 30.4 & 27.0 \\
 6 & 0.48 & 50 &    & 28 & 20 & 39.4 &    & 32.0 & 29.2 \\
 7 & 0.36 & 40 &    & 22 & 20 & 42.4 &    & 36.0 & 33.4\\
 8 & 0.30 & 40 &    & 22 & 22 & 44.4 &    & 38.2 & 37.8 \\
 9 & 0.26 & 40 &    & 22 & 22 & 46.4 &    & 38.0 & 38.8 \\
10 & 0.23 & 40 &    & 36 & 36 & 49.4 &    & 46.6 & 47.0 \\
11 & 0.20 & 40 & 52 & 36 & 28 & 53.8 & 57 & 48.4 & 43.8 \\
\botrule
\end{tabular}}
\begin{tabnote}
$^{\text a}$ The numbers shown in this column interpolate the
  results presented in Ref.~\cite{baldo}.\\
\end{tabnote}
\label{jm:tab1}
\end{table}
The inner crust matter is divided in 11 zones shown in
Tab.~\ref{jm:tab1}, as in Ref.~\cite{negele} (notice that the denser
zone has been removed since it is in the deformed pasta region).
Each W-S cell is supposed to contain in its center a spherical
neutron-rich nucleus surrounded by unbound neutrons and immersed in a
relativistic electron gas uniformly distributed inside the cell.
For a given baryonic density $\rho_B=A/V$, neutron number $N$ and
proton number $Z$, the total energy of a W-S cell, $E_{tot}$ has
contributions coming from the rest mass of the particles, 
$E_m(N,Z)=Z(m_p+m_e)+Nm_n$, 
from the nuclear components (including Coulomb interaction between protons) 
$E_n(\rho_B,N,Z)$, 
from the lattice energy, 
\be
E_l(\rho_B,N,Z,\{\rho_p\})=\frac{1}{2}\int d^3r_1 d^3r_2
\rho_e(r_1)\frac{e^2}{|r_{12}|} \left(\rho_e(r_2)-\rho_p(r_2)\right) \; ,
\ee
which is induced by the difference between the uniform electron density and 
proton density, and finaly from the kinetic energy $T_e$ of the relativistic electron
gas~\cite{negele}. The Coulomb exchange energy as well as the
screening correction for the electrons gas has been neglected.

The self-consistent Hartree-Fock-Bogoliubov (HFB) approach has been presented in
detail in Ref.~\cite{doba}. It has been extended to describe the
dense part of the inner crust at finite temperature in
Ref.~\cite{sandulescu2}. With minor modifications, it has been
successfully applied to calculate the specific heat of the whole
part of the inner crust~\cite{mon07}. The 
effective N-N interaction should provide a reasonable description
of both the nuclear clusters and the neutron gas. We use the
Skyrme force SLy4~\cite{sly4} which was 
adjusted to describe properly the mean field properties of
neutron-rich nuclei and infinite neutron matter. The choice of the
pairing force is more problematic since at present it is not yet
clear what is the strength of pairing correlations in low density neutron
matter. However, in the present paper we have chosen to adjust the
effective pairing interaction on the results of the pairing gap
obtained with the bare interaction. We then have a maximum gap in
neutron matter of about 3~MeV located at $k_F=0.85$~fm$^{-1}$.

The minimization is performed in two steps: first, at a given
density $\rho_B$ and for a given mass number $A$, we look for the
cell which satisfies the
beta-stability criterion, $\mu_p+\mu_e=\mu_n$. The neutron and
proton chemical potentials are provided by the HFB
calculation while the electron chemical potential is deduced from
$\mu_e=dE_{tot}/dN_e$. Note that,
due to shell effects we may find several cells which satisfy this
criterion. We then choose the one which has a minimal energy. In
the second step, we minimize the total energy $B_{tot}$ with
respect to the mass number $A$. It should be noted that, in the
second step all the cells must be calculated at the same density
while the minimum energy is searched over the variable $A$. 
As the volume is modified by unit steps (the step of the mesh is
0.2~fm), the mass number $A$ can take only discrete values so that
$A=\rho_B V$ is always satisfied. The step between two values of
$A$ is not constant and vary as the square of the WS radius
$R_{WS}$. We have approximatively $dA\sim\rho_B 4\pi
R_{WS}^2dR_{WS}$.

\begin{figure}[t]
\begin{center}
\psfig{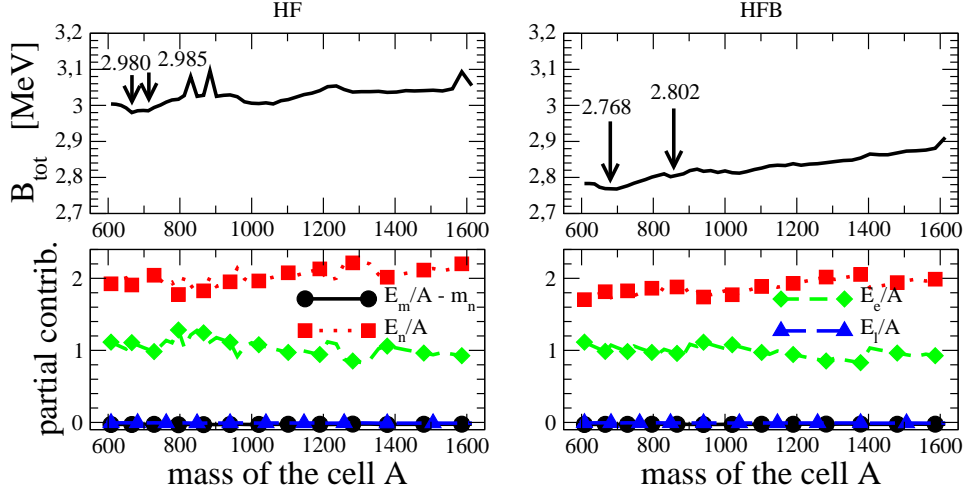}
\end{center}
\caption{
The binding energy $\epsilon/\rho_B$ of the
  W-S cells in zone $N_{zone}$=4 as well as the partial
  contributions of the rest mass energy $E_m/A-m_n$, the nucleons, the
  electrons and the lattice. 
  HF and HFB 
  results are shown on the left and right panels, respectively. 
See the text for more details.}
\label{jm:fig1}
\end{figure}

For $N_{zone}$=4, we show in Fig.~\ref{jm:fig1} a typical example
of step 2 in the search for the W-S cell which minimizes the
energy. On the l.h.s. is represented the total binding energy
$E_{tot}/A$ of the W-S cells obtained in Hartree-Fock (HF)
mean-field model (without pairing). The position of the lowest
minimum is indicated as well as two other local minima. In this
case, there is a difference of 5~keV per nucleon between the two
first minimum. On the r.h.s., we plot the results of HFB
calculations which show the effects of the pairing correlations:
the first minimum is 
deeper (about 34~keV per nucleons). On the lower part of the graph, we show the partial
contributions of the rest mass energy $E_m/A-m_n$, the nuclear binding
energy $E_n/A$, the electron kinetic energy $T_e/A$ and lattice energie $E_l/A$. 
One sees that the inner crust is a typical example of frustrated
system: when the nuclear energy is minimum, the electron energy is
maximum and vice-versa. The minimum in energy is a compromise
between these two opposite energies. 
The consequence is that it gives rise to numerous local minima which
are not very far in energy from the absolute minimum.

The equation of state obtained at the HF and HFB approximations is
given in Table~\ref{jm:tab1}. In total, 19714 cells configurations has
been calculated. The pairing correlations do not
strongly modify the HF results, it may slightly decrease the
proton number and the WS radius. We also show
previous calculations of equations of state. We have obtained
proton numbers $Z$ different from those of
Refs.~\cite{negele,baldo}. The results of Ref.~\cite{negele} have
been obtained in a density matrix expansion framework
interpolating microscopic calculations for homogeneous matter. The
difference may be due to the fact that the Skyrme interaction SLy4
is based on more recent microscopic parametrizations. It is
difficult to compare with the results of Ref.~\cite{baldo} since
those calculations have not been done at the same densities and
they have essentially been carried out in the dense part of the
crust. We then show an interpolation of their results. The
comparison 
shows a similar trend in the dense cells.

\section{Discussion of the shell effects}

In ordinary nuclei, the bound states play an essential role in
determining the ground state properties and the role of
the continuum states is usually negligeable. The coupling between
bound states and continuum states can be more important for nuclei
near the drip lines, as the Fermi energy is increasing. However,
even in these most extreme situations continuum states are only
virtually populated and the occupation numbers associated to
continuum states are small compared to those of the bound states.
In the inner crust of neutron stars, the situation is completely
different: the Fermi energy is positive and most of the mass is
due to the unbound neutrons while bound states account only for a
small fraction. 
Thus, the continuum states contribute 
in an essential way to physical properties of the crust like
its density, total energy or specific heat. The presence of bound
states, or nuclear clusters, is not negligible but it induces only
corrections to the ground-state properties~\cite{mon07}. 
It is then very important to have an accurate model of the unbound
states.

It should also be noted that, as the continuum states are embedded
in a Coulomb lattice, their asymptotic behavior is different from
that occurring in nuclei. The continuum states are indeed
periodically distorted within a distance equal to the lattice step
(20-100~fm). The proper theory for the description of the
continuum states is the band theory which requires to solve the
Schr\"odinger equation~\cite{chamel,cha07}, $(h_0+h_{{\bf
k}})u_{\alpha,{\bf k}}({\bf r})=\epsilon_{\alpha,{\bf k}}
u_{\alpha,{\bf k}}({\bf r})$, where $h_0$ is the usual
single-particle Hamiltonian of mean-field models. The second term
$h_{{\bf k}}$ is induced by the Floquet-Bloch boundary conditions
for the single-particle wave function $\varphi_{\alpha,{\bf k}}$:
$\varphi_{\alpha,{\bf k}}({\bf r})=u_{\alpha,{\bf k}}({\bf r})
e^{i{\bf k}\cdot{\bf r}}$. The spherical W-S approximation is
obtained by setting ${\bf k}=0$ and replacing the W-S elementary
polyhedron by a sphere. Then, the remaining index $\alpha$ is
discrete, like in usual mean-field models, and runs over the
eigenstates of the Hamiltonian $h_0$. The main difference between
mean-field models and band theory is indeed due to the term
$h_{{\bf k}}$. In band theory, this term introduces a new index
${\bf k}$ which is a continuous variable ranging between 0 and
$\pi/a$~fm$^{-1}$ where $a$ is the lattice step. As a consequence,
the density of unbound states is continuous (notice that it may
also have structures and gaps). 
It is then clear that the discrete distribution
of unbound states present in usual mean-field models is a
consequence of the dropping of $h_{{\bf k}}$ in the W-S
approximation.

We can characterize the distribution of unbound states by the
average distance between the unbound states energies, which can be
related to the radius of the W-S cell as $\Delta
\epsilon\sim\hbar/2mR_{WS}^2$: the smaller the W-S radius, the
larger the average distance $\Delta\epsilon$. One could notice
that the W-S radius is a function of the density and it is fixed
by the equation of state, as shown in Table~\ref{jm:tab1}. At the
W-S approximation, the distribution of unbound states is then also
fixed by the density. The non-continuous distribution of unbound
states could be interpreted as a spurious effect, induced by the
W-S approximation, since in band theory calculations, the density
of unbound states varies in a continuous way~\cite{cha07}. One can
then distinguish between the physical shell effects due to the
bound states from the spurious shell effects due to the unbound states
and the W-S approximation.

\section{Spurious shell effects}
\label{jm:sec2}

In order to estimate the spurious shell effects due to the unbound
neutrons, we simulate an homogeneous gas of neutron matter in W-S
cell by removing all the proton states. In homogeneous matter, the
energy density should be independent of the volume, then of the
W-S radius, and it should depend only on the density. In our
simulation of homogeneous matter, we then vary the W-S radius from
10 to 50~fm, at constant neutron density. We call
$B_{WS-hom.}(\rho_{unb.},R_{WS})$ the binding energy per particle
obtained for a given W-S radius $R_{WS}$ and for a given density
of unbound neutrons $\rho_{unb.}$.
\begin{figure}[t]
\begin{center}
\psfig{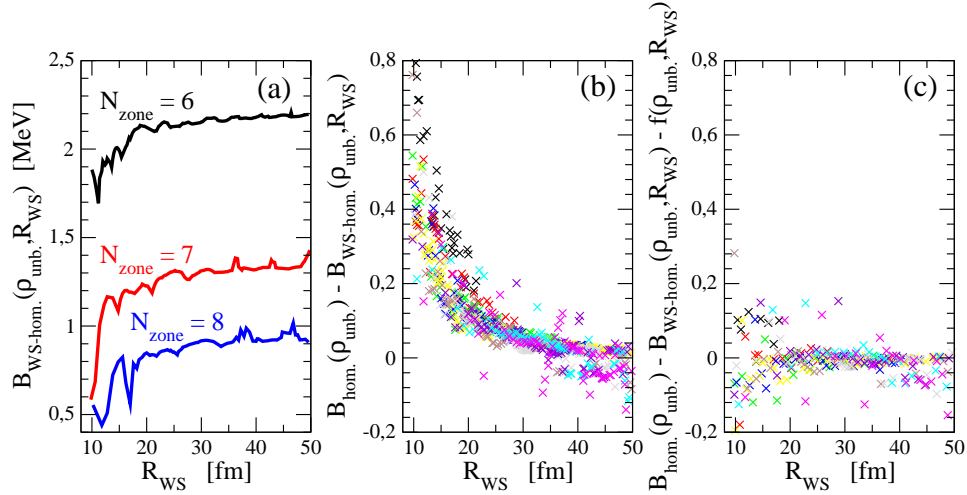}
\end{center}
\caption{We represent several quantities versus the W-S radius:
(a) the binding energy $B_{WS-hom.}(\rho_{unb.},R_{WS})$ corresponding to
$N_{zone}$=6, 7 and 8
(b) the difference
$B_{hom.}(\rho_{unb.})-B_{WS-hom.}(\rho_{unb.},R_{WS})$ (dots) and
the function $f(\rho_{unb.},R_{WS})$ which fit the smooth component
(solid lines), and
(c) the difference between the dots presented in (b) and the function
$f(\rho_{unb.},R_{WS})$.}
\label{jm:fig2}
\end{figure}
We show the binding energy versus the W-S radius in
Fig.~\ref{jm:fig2}(a) for several densities corresponding to
$N_{zone}$=6, 7 and 8. It is clearly seen that the binding energy
is not independent of the W-S radius, except for large W-S radii
where the binding energy converge to the energy of homogeneous
matter, $B_{hom.}(\rho_{unb.})$. The binding energy of homogeneous
matter $B_{hom.}(\rho_{unb.})$ is calculated from the average
value of the Skyrme two-body interaction on a plane wave basis.
The difference
$B_{hom.}(\rho_{unb.})-B_{WS-hom.}(\rho_{unb.},R_{WS})$ is
represented on Fig.~\ref{jm:fig2}(b) (dots). As expected, this
quantity is converging to zero for large W-S radii, but for small
radii, around 20~fm for instance, the difference can be as large
as 300~keV per nucleons.

From the dots represented on Fig.~\ref{jm:fig2}(b), we obtain an
universal function which fits the smooth radial dependence of the
difference
$B_{hom.}(\rho_{unb.})-B_{WS-hom.}(\rho_{unb.},R_{WS})$. As this
effect is due to the discretization of the unbound states, it is
proportional to the average distance $\Delta\epsilon$ between the
box states. 
This difference should then decrease with the W-S radius as
$R_{WS}^{-2}$. It should also be proportional to the number of
unbound neutrons, while the density dependence is not known. We
then represent it by a power law $\rho_{unb.}^\alpha$ where the
power $\alpha$ is adjusted to reproduce the dots presented in
Fig.~\ref{jm:fig2}(b). 
In this figure we show the fitted function, 
\be
f(\rho_{unb.},R_{WS}) = 89.05 (\rho_{unb.}/\rho_0)^{0.1425}
R_{WS}^{-2} \; . 
\ee 
It interpolates the smooth part of the shell effects on the
binding energy. There is, however, a residual difference which
cannot be fitted. Finally, to estimate the residual difference
between 
$B_{hom.}(\rho_{unb.})-B_{WS-hom.}(\rho_{unb.},R_{WS})$ and the
adjusted function $f(\rho_{unb.},R_{WS})$, shown on the r.h.s of
Fig.~\ref{jm:fig2}(c). 
The average residual fluctuations are now of the order of 50~keV.

\section{Modified HF binding energy in W-S cells (HF-WS)}
\label{jm:sec3}

In Sec.~\ref{jm:sec2}, it has been shown that the fluctuations in
the binding energy induced by the spurious shell effect can be
decomposed into a smooth and a residual term and we have obtained a
universal function for the smooth term.
We then propose a systematic correction to the HF binding energy
in the W-S cell: 
\be 
B_{HF-WS}(\rho_B,\rho_{unb.},R_{WS}) &\equiv&
B_{WS-inhom.}(\rho_B,R_{WS}) \nonumber \\
&&+\Big(B_{hom.}(\rho_{unb.})-B_{WS-hom.}(\rho_{unb.},R_{WS}) 
\Big)_{smooth}
\nonumber \\
&=& B_{WS-inhom.}(\rho_B,R_{WS})+f(\rho_{unb.},R_{WS})
\; . 
\ee 
In the calculation of the total energy, the nuclear binding energy $E_n/A$
should then be replaced by $B_{HF-WS}(\rho_B,\rho_{unb.},R_{WS})$.
The minimum energy can be unambiguously identified if the difference
in energy between the first and the second minimum is larger that the
residual fluctuation, about 50~keV. 
In the HF calculation, the difference in energy between different cell 
configurations are of the order of few keV to few tens of keV.
The residual fluctuation of the proposed correction is then too
large.
However, as seen above, the pairing correlations help in producing
a deeper first minimum. It may also help in reducing the size of
the residual fluctuations. 
In a future investigation, we 
plan to obtain an improved binding energy for W-S cells including pairing
correlations. It must be remarked that the effects of the coupling between 
the protons and the unbound neutrons have been neglected in $B_{HF-WS}$. 
This is a good approximation since the protons are deeply bound.



\section{Conclusions}

In the search for the W-S cells which minimize the energy, it is
important to look not only at the lowest minimum energy, but also
at the other minima. The differences 
between the first and other minima 
show how stable is the lowest minimum with respect, for instance,
to thermal fluctuations or spurious shell effects. In HF
calculations, the differences in total energy between the first
minimum and the others are about few keV per nucleon (few
MeV in total energy). It 
is shown that the spurious shell effects, induced by the W-S
approximation, introduce a fluctuation
in the calculation of the total energy up to about 300~keV per
nucleon. Interpolating the smooth part of the spurious shell
effects, we have then proposed a simple method to reduce the
fluctuations down to about 50~keV per nucleon (HF-WS). 
The residual fluctuations are still too large, and the removal of
the smooth part is not enough to improve the results 
obtained within the W-S approximation.
We have also
shown that the pairing correlations help in producing a deeper
first minimum. We have then compared the equation of state
obtained from HF and HFB binding energies.
Several issues should still be addressed in future studies: 
spurious shell correction with pairing correlations as well as
effects of the temperature. These two effects should give lower
residual fluctuations than the one we obtained at the $T=0$ HF 
approximation.

\end{document}